\documentclass[10pt,journal,twocolumn]{IEEEtran}
\usepackage{fixltx2e}
\usepackage{verbatim}
\usepackage{hyperref}
\usepackage[cmex10]{amsmath}
\usepackage{graphicx}
\usepackage[center]{caption2}
\usepackage{stfloats}
\usepackage{algorithm,algpseudocode}
\usepackage{cases}
\usepackage{array}
\usepackage{setspace}
\usepackage{fancyhdr}
\usepackage{amscd}
\usepackage{enumerate}
\usepackage{multirow}
\usepackage{float}
\usepackage{color}
\usepackage{longtable}
\usepackage{algorithm}
\usepackage{epstopdf}
\usepackage{pgf}
\usepackage{amsmath}
\usepackage{mdwmath}
\usepackage{mdwtab}
\usepackage{float}

\usepackage{multicol}
\usepackage{lipsum}

\usepackage[caption=false,font=footnotesize]{subfig}

\usepackage{fixltx2e}
\usepackage{epstopdf}
\usepackage{amsthm}
\usepackage{subfig}
\usepackage{hyperref}

\begin{document}
	
\title{Joint User-Association and Resource-Allocation in Virtualized Wireless Networks}

\author{
	\IEEEauthorblockN{ Saeedeh Parsaeefard\IEEEauthorrefmark{1}, Rajesh Dawadi\IEEEauthorrefmark{1}, Mahsa Derakhshani\IEEEauthorrefmark{2}, Tho Le-Ngoc\IEEEauthorrefmark{1}}\\
	\IEEEauthorblockA{\IEEEauthorrefmark{1}Department of Electrical \& Computer Engineering, McGill University, Montreal, QC, Canada}
	\IEEEauthorblockA{\IEEEauthorrefmark{2}Department of Electrical \& Computer Engineering, University of Toronto, Toronto, ON, Canada
		\\Email:  Saeideh.parsaeifard; rajesh.dawadi@mail.mcgill.ca; \\mahsa.derakhshani@utoronto.ca; tho.le-ngoc@mcgill.ca }
}	
\vspace{-10mm}
\maketitle
\vspace{-15mm}
\begin{abstract}
	
In this paper, we consider a down-link transmission of multicell virtualized wireless networks (VWNs) where users of different service providers (slices) within a specific region are served by a set of base stations (BSs) through orthogonal frequency division multiple access (OFDMA). In particular, we develop a joint BS assignment, sub-carrier and power allocation algorithm to maximize the network throughput, while satisfying the minimum required rate of each slice. Under the assumption that each user at each transmission instance can connect to no more than one BS, we introduce the user-association factor (UAF) to represent the joint sub-carrier and BS assignment as the optimization variable vector in the mathematical problem formulation. Sub-carrier reuse is allowed in different cells, but not within one cell. As the proposed optimization problem is inherently non-convex and NP-hard, by applying the successive convex approximation (SCA) and complementary geometric programming (CGP), we develop an efficient two-step iterative approach with low computational complexity to solve the proposed problem. For a given power-allocation, Step 1 derives the optimum user-association and subsequently, for an obtained user-association, Step 2 find the optimum power-allocation. Simulation results demonstrate that the proposed iterative algorithm outperforms the traditional approach in which each user is assigned to the BS with the largest average value of signal strength, and then, joint sub-carrier and power allocation is obtained for the assigned users of each cell. Especially, for the cell-edge users, simulation results reveal a coverage improvement up to $57\%$ and $71\%$ for uniform and non-uniform users distribution, respectively leading to more reliable transmission and higher spectrum efficiency for VWN.  

\end{abstract}
\section{Introduction}	
\subsection{Motivation}

The next generation of wireless networks encounter different types of challenges such as meeting the requirements of new high-data rate applications, reducing the cost, and increasing the spectrum and infrastructure efficiency. In this context, recently, virtualization of wireless networks has been brought up as a promising approach to tackle these issues. In \textit{virtualized wireless networks} (VWNs), the physical resources of one network provider, e.g., spectrum, power, and infrastructure, are shared between different service providers, also called slices \cite{6117098,6887287,zaki2010lte}. Generally, each slice comprises of a set of users, and has its own quality-of-service (QoS) requirements. To harvest the potential advantages of VWN, effective resource allocation is a major concern which has been drawing a lot of attention in recent years.

For instance, in \cite{6117098}, a resource management scheme is studied by introducing two types of slices, including rate-based and resource-based slices, where the minimum rate and minimum network resources are preserved for each slice, respectively. Furthermore, in \cite{6177700}, interactions among slices, network operator, and users are modeled as an auction
game where the network operator is responsible for spectrum
management on a higher level, and each slice focuses on QoS
management for its own users. To preserve the QoS of slices from wireless channel fading, the admission control policy is proposed in \cite{saeedeh-vikas-WCNC} where the requirement of each slice is adjusted by the channel state information (CSI) of its own users. To extend the feasibility condition of VWN in order to support diverse QoS requirements, \cite{jumbaresource} considers the use of massive MIMO where the access point is equipped with a large number of antennas. In \cite{6314178}, the combination of time, space and elastic resource allocation for OFDMA system is considered. Advantages of full-duplex transmission relay in VWN have been investigated in \cite{7037591}.

Generally, these works have focused on analyzing the resource allocation problem in a single-cell VWN scenario. However, in practice, the coverage of a specific region may require a set of BS, in a multi-cell VWN scenario. The key question in a multi-cell VWN scenario is how to allocate the resources to maintain the QoS of each slice, while improving the total performance of VWN over a specific region. In this paper, we consider a multi-cell OFDM based VWN where the coverage of a specific region is provided by a set of BSs serving different groups of users belonging to different slices. The QoS of each slice is represented by its minimum reserved rate. Each user of each slice can be only served by one BS and this BS is not predetermined by the distance or by measuring the average received signal strength of BSs. Consequently, in this setup, the resource sets in the related optimization problem involve the \textit{set of BSs, sub-carriers and power} for each user belonging to each slice.  

In this paper, the objective of proposed resource allocation problem is to maximize the total throughput of VWN in the specific region subject to power limitation of BSs, minimum required rate of each slice, and sub-carrier and BS assignment limitations. Based on the limitations of down-link OFDMA transmission, each sub-carrier can be assigned to one user within a cell and each user can be associated to only one BS. Since in this optimization problem, the sub-carrier assignment and BS association are inter-related, we introduce the user association factor (UAF) that jointly determines the BS assignment and sub-carrier allocation as the optimization variable vector. Due to this user-association constraint and the  inter-cell interference, the proposed optimization problem is non-convex and NP-hard, suffering from high computational complexity \cite{4453890}. We apply the frameworks of complementary geometric programming (CGP) and the successive convex approximation (SCA) approach \cite{GPmungchiang,4275017,xu2014global,Derakhshani2015full} to develop an efficient, iterative, two-step algorithm to solve the proposed problem. For a given power-allocation, Step 1 derives the optimum user-association solution, and subsequently, with this obtained user-association solution, Step 2 finds the optimal  power allocation. Such 2-step optimization process is repeated until convergence. Furthermore, it can be shown that even the simplified problem of each step still involves non-convex optimization problem. Thus, by applying various transformation and convexification techniques, we develop the analytical framework to  transform  the non-convex optimization problems encountered in each step into the equivalent lower-bound geometric programming (GP) problems, \cite{xu2014global}, which can be solved by efficient softwares, e.g., CVX \cite{cvx}.      

Simulation results demonstrate that the proposed approach can significantly outperform the traditional scenario where the BS assignment is based on the largest average signal-to-interference-plus-noise ratio (SINR), and subsequent sub-carrier and power allocation is derived for the users of each cell. The simulation results reveal that considering UAF can increase the feasibility of resource allocation problem (the required rate of each slice will be satisfied with a higher probability as compared to the traditional approach). Specifically, the proposed algorithm can significantly increase the probability of achieving higher rates for  the cell-edge users, resulting better coverage for the VWN.

\subsection{Related Works}
Our work in this paper lies along the intersection of two research contexts in resource allocation problems: 1) multi-cell OFDMA wireless networks, and 2) VWNs.

There exists a large body of research conducted in resource allocation for multi-cell OFDMA wireless networks. For example, in \cite{kim2013sum}, the resource allocation in conventional OFDMA-based network is studied using BS-assignment based on the largest average received signal\footnote{This average is derived based on the measurement of users over one specific window in both idle and active phases where the size of window and the measurement of each user are adjusted based on specification of wireless network standards \cite{goldsmithbook}. } strength from the BS at each user. An iterative algorithm for maximizing the weighted sum of minimal user rates in each BS is explored in \cite{wang2011iterative}. Joint cell, channel and power allocation in multi-cell relay networks is explored in \cite{fallgren2012optimization}, where each user is assigned to the BS with the highest channel gain. In \cite{chohee}, a proportional fair resource allocation in a multi-cell OFDMA networks is proposed with the aim of maintaining the quality of experience of users by considering a utility function based on the mean opinion score. In \cite{yu2014multi}, joint scheduling of resource blocks, power allocation, and modulation and coding scheme in LTE-A system is considered by using the criteria of proportional fairness. A similar problem in OFDMA cognitive radio networks is studied in \cite{forouzan2014new} where an iterative algorithm is proposed to solve the sub-carrier and power allocation. Similarly, in \cite{Wenpeng}, a resource allocation problem for jointly optimizing the energy and spectral efficiency is proposed for a multi-cell OFDMA wireless network by considering an energy and spectral efficiency trade-off metric. In \cite{wangEE}, the authors have considered an energy efficient resource allocation problem for a multi-cell OFDMA network in a conventional wireless network where the available values of channel state information (CSI) are imperfect. In \cite{bianchioptimal}, a resource allocation algorithm is proposed for a two-cell down-link OFDMA network with a fractional frequency reuse scheme between BSs. 

In the aforementioned works (i.e., \cite{kim2013sum,wang2011iterative,fallgren2012optimization,chohee,yu2014multi,forouzan2014new,Wenpeng,bianchioptimal}), the BS assignment algorithm is separated from the sub-carrier allocation, while joint sub-carrier and power allocation is applied for multi-cell scenario. Compared to this research direction, we consider UAF which jointly assigns the BS and sub-carrier for each user. In our proposed algorithm, after assigning UAF, the power allocation algorithm is applied. Furthermore, we maintain the implementation limitations of multi-cell OFDMA networks by proposing new constraints.

As previously mentioned, the resource allocation in VWNs has received growing attention. In \cite{6887287}, different aspects of VWN including resource discovery and allocation as well as the research challenges have been discussed. Besides \cite{6117098,6177700,saeedeh-vikas-WCNC,jumbaresource,6314178}, in \cite{6966044}, the challenge in allocating physical resource blocks (PRBs) to various slices in an LTE network has been addressed considering a single BS scenario. In \cite{yang2014opportunistic}, an opportunistic algorithm to allocate the resources to virtual operators is proposed by differentiating the resource requirements among operators as baseline and fluctuate requirements to ensures the minimum QoS requirements of each virtual operator. In \cite{zaki2010lte}, the concept of virtualization has been extended to a LTE network by considering virtual operators or slices each with various bandwidth requirements in terms of the physical resource blocks (PRBs) in LTE. To the best of our knowledge, the multi-cell scenario of VWNs has not been studied in the previous related works in this context. 

\subsection{ Structure of Paper}

The rest of this paper is organized as follows. In Section II, the system model and problem formulations are introduced. Section III contains mathematical background, the detailed explanation of iterative algorithms and computational complexity analysis. Section IV demonstrates the simulation results and their detailed analysis followed by concluding remarks in Section V.

\vspace{-5mm}
\section{System Model}

\begin{figure}[!t]
	\centering
	\includegraphics[width=1.4\linewidth]{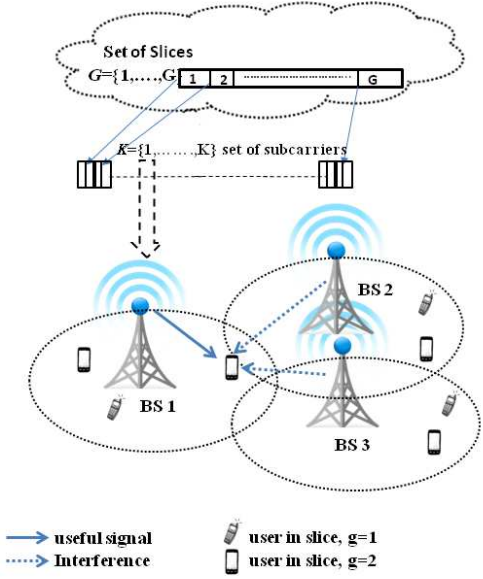}
	\caption{Illustration of multi-cell VWN with $M$ BSs serving sets of users belonged to different slices.} 
	\label{Model Pic}
\end{figure}

We consider the down-link transmission of a VWN where the coverage of a specific area is provided by the set of BSs, i.e., $\mathcal{M}= \{1,\dots,M\}$. The total bandwidth of $B$ Hz is divided into a set of sub-carriers, $\mathcal{K}= \{1,\dots,K\}$ and shared by all BSs through orthogonal frequency-division multiple-access (OFDMA). The bandwidth of each sub-carrier, i.e., $B_c =\frac{B}{K}$, is assumed to be much less than the coherent bandwidth of the wireless channel, meaning that the channel response in each sub-carrier is flat. This set of BSs serves a set of slices, i.e., $\mathcal{G}= \{1,\dots,G\}$, where the slice $g$ has a set of  users (denoted by $\mathcal{N}_{g}= \{1,\dots,N_{g}\}$) and requests for a minimum reserved rate of $R_g^\text{rsv}$. A typical illustration of this setup is shown in Fig. \ref{Model Pic}, where $N=\sum_{g\in \mathcal{G}}N_g$ is the total number of users.

Let $h_{m,k,n_g}$ and $P_{m,k,n_g}$ be the channel state information (CSI) of the link from BS $m \in \mathcal{M}$ to user $n_g$ of slice $g$ on sub-carrier $k$ and allocated power of user $n_g$ of slice $g$ on sub-carrier $k$, respectively. In this scenario, due to the OFDMA limitation, each user is assigned to one BS, and to avoid intra-cell interference, orthogonal sub-carrier assignment is assumed among users in a cell. We assume that there is no pre-allocated BS for users, and consider $\beta_{m,k,n_g} \in \{0,1\}$, which represents both sub-carrier allocation and BS assignment indicator for user $n_g$ of slice $g$ on sub-carrier $k$ of BS $m$ which has been called user association factor \textit{(UAF)}. Note that $\beta_{m,k,n_g}=1$ when BS $m$ allocates sub-carrier $k$ to user $n_g$, and $\beta_{m,k,n_g}=0$, otherwise. 

Consider $\textbf{P}=\left[P_{m,k,n_g}\right]_{\forall m, g, n_g, k}$ and $\boldsymbol{\beta}=\left[\beta_{m,k,n_g}\right]_{\forall m, g, n_g, k}$ as the vector of all transmit powers and UAFs of users, respectively. Therefore, the rate of user $n_g$ at sub-carrier $k$ of BS $m$ can be expressed as
\begin{equation}
\label{eq:rate_eqn}
R_{m,k,n_g}{(\textbf{P})}=\log_2\left[1+\frac{P_{m,k,n_g}h_{m,k,n_g}}{\sigma^2+I_{m,k,n_g}}\right],
\end{equation}
where $I_{m,k,n_g}=\\ \sum_{\forall{m'}\in\mathcal{M}, m'\neq{m}}\sum_{\forall{g}\in\mathcal{G}}\sum_{\forall{n_g'}\in\mathcal{N}_g, n_g'\neq{n_g}}P_{m',k,n'_g}h_{m, k,n_{g}'}$ is the interference observed by user $n_g$ in cell $m$ and sub-carrier $k$, and $\sigma^{2}$ is the noise power. Without loss of generality, noise power is assumed to be equal for all users in all sub-carriers and BSs. Now, from \eqref{eq:rate_eqn}, the required minimum rate of slice $g \in \mathcal{G}$ can be represented as
\begin{align*}\label{C1}
\text{C1}:  \sum_{m\in \mathcal{M}}\sum_{n_g\in\mathcal{N}_g}\sum_{k\in \mathcal{K}}{\beta_{m,k,n_g}}R_{m,k,n_g}(\textbf{P})\geq{R_g^\text{rsv}}, \forall g \in \mathcal{G}.
\end{align*}
Considering the maximum transmit power limitation of each BS, 
\begin{equation*}\label{C2}
\text{C2}: \quad \quad\quad \sum_{g\in{\mathcal{G}}}\sum_{n_g\in\mathcal{N}_g}\sum_{k\in{K}}P_{m,k,n_g} \leq{P_m^\text{max}}, \quad\forall{m}\in \mathcal{M},
\end{equation*}
where $P_m^{\text{max}}$ is the maximum transmit power of BS $m$. Furthermore, the OFDMA exclusive sub-carrier allocation within each cell $m$ can be expressed as
\begin{equation*}\label{C3}
\text{C3}:\quad \quad \quad \sum_{g\in{\mathcal{G}}}\sum_{n_g\in\mathcal{N}_g}\beta_{m,k,n_g}\leq1,\quad \quad  \forall{m}\in{\mathcal{M}}, \quad  \forall{k}\in{\mathcal{K}}.
\end{equation*}
In this setup, due to the limitation of multi-cell OFDMA, we restrict the access of each user by the following constraint
\begin{equation*}\label{C4}
\begin{split}
\text{C4}:\quad & \big[\sum_{k\in\mathcal{K}}\beta_{m,k,n_g}\big]\big[\sum_{\forall{m'}\neq{m}}\sum_{k\in\mathcal{K}}\beta_{m',k,n_g}\big]=0,\quad \\
& \forall{n_g\in \mathcal{N}_g},\,\,\forall{g}\in \mathcal{G},\,\, \forall{m} \in \mathcal{M}.
\end{split}
\end{equation*}
C4 imposes that when sub-carrier $k$ from BS $m$ is assigned to user $n_g$, it will not be assigned to user $ n_g $ by BS $m'$ where $m'\neq m$ and  $\forall m', m \in \mathcal{M}$. C4 is novel since it jointly assigns the BS and sub-carrier over the specific region. Therefore, we call $\beta_{m,k,n_g}$ as the user association factor (UAF) which simultaneously determines the BS and sub-carrier in the VWN. The joint power, sub-carrier and BS assignment can be formulated as
\begin{align} \label{original_problem}
& \max_{\boldsymbol{\beta},\textbf{ P}}\sum_{m\in \mathcal{M}}\sum_{g\in \mathcal{G}}\sum_{n_g\in\mathcal{N}_g}\sum_{k\in\mathcal{K}}\beta_{m,k,n_g}R_{m,k,n_g}(\textbf{P}),   \\
& \text{subject to: C1 - C4.} \nonumber
\end{align}
The optimization problem introduced in \eqref{original_problem} has a non-convex objective function due to inter-cell interference and involves non-linear constraints with combination of continuous and binary variables, i.e., $\textbf{P}$ and $\boldsymbol{\beta}$. In other words, \eqref{original_problem} is a non-convex mixed-integer, NP-hard optimization problem \cite{5770666,4453890}. Therefore, proposing the efficient algorithm with reasonable computational complexity is desirable. 

\section{Two-Level Iterative Algorithm for Joint User Association and Resource Allocation}

To tackle the computational complexity of \eqref{original_problem}, we adopt an iterative approach to iteratively find the UAF and power allocation for each user in two steps as shown in Algorithm 1. In Step 1, for given power allocation vector, the UAF considered as the variable of the user-association problem is solved by Algorithm 1.A to be discussed in detail in Section III.B. 

This derived UAF is then used in Step 2 to find the corresponding allocated power as the solution of the power-allocation optimization problem by Algorithm 1.B to be discussed in detail in Section II.C. Steps 1 and 2 are iteratively executed until both the current UAF and power allocation vector solutions are not much different from their values obtained in the previous iteration. In other words, the sequence of the UAF and power allocation vector solutions can be expressed as
\begin{equation}
\underbrace{\boldsymbol{\beta}(0)\rightarrow\textbf{P}(0)}_\text{Initialization}\rightarrow\dots\underbrace{\boldsymbol{\beta}(t)^{*}\rightarrow\textbf{P}(t)^{*}}_\text{Iteration $t$}\rightarrow\underbrace{\boldsymbol{\beta}^{*}\rightarrow\textbf{P}^{*}}_\text{Optimal solution},
\end{equation}
where $t>0$ is the iteration number and $\boldsymbol{\beta}(t)^{*}$ and $\textbf{P}(t)^{*}$ are the optimal values at the iteration $t$ from convex transformation of related optimization problems in each step. The iterative procedure is stopped when
\begin{equation*}
\vert\vert{\boldsymbol{\beta}}^*(t)-\boldsymbol{\beta}^*(t-1)\vert\vert\leq{\varepsilon}_1 \text{ and }
\vert\vert{\textbf{P}}^*(t)-\textbf{P}^*(t-1)\vert\vert\leq{\varepsilon}_2
\end{equation*}
where $0 < \varepsilon_{1} \ll 1$ and $0<\varepsilon_{2} \ll 1$. 

Notably, both the user-association and power-allocation optimization problems are still non-convex and suffer high computational complexity. To solve them efficiently, we apply complementary geometric programming (CGP) for each step \cite{xu2014global} in which via different transformations and convexification approaches, the sequence of lower bound GP based approximation of relative optimization problem is solved as described in detail in the following sections.

\subsection{Complementary geometric programming (CGP):A brief review}

\begin{algorithm}[t]
	\caption{Iterative Joint User Association and Power Allocation Algorithm}
	\label{Algorithm1}
	\textbf{Initialization:} Set $t:=1$, $\boldsymbol{\beta}(t=1)=[\textbf{1}]$, where \textbf{1} is a vector, $C^{1\times KN}$ and the power over of each sub-carrier of BS $m$ is equal to ${P_{m}^{\text{max}}}/{K}$.\\
	\textbf{Repeat} 
	\\
	
	\hspace{.5cm}\textbf{Step 1:} Derive $\boldsymbol{\beta}^*(t)$ from \eqref{beta_eqn} by using Algorithm \textbf{1.A};\\
	
	\hspace{0.5cm}\textbf{Step 2:} For derived $\boldsymbol{\beta}^*(t)$, find $\textbf{p}^*(t)$ from \eqref{power_eqn} by applying Algorithm \textbf{1.B};
	\\
	
	\hspace{0.5cm}\textbf{Step 3:} \textbf{Stop} if $\vert\vert{\boldsymbol{\beta}}^*(t)-\boldsymbol{\beta}^*(t-1)\vert\vert\leq{\varepsilon_1}$, and $\vert\vert{\textbf{p}}^*(t)-\textbf{p}^*(t-1)\vert\vert\leq{\varepsilon_2}$, otherwise, set $t:=t+1$ and go to \textbf{Step 1}.
\end{algorithm}

\label{background}
Geometric programming (GP) is a class of non-linear optimization problems which can be solved very efficiently via numerical methods \cite{4275017}. Thus, a significant amount of research has been done in order to convert the resource allocation problems into GP problems, so that it becomes computationally tractable \cite{GPmungchiang,4275017,Boyd222,965961,1258939}. 

The standard form of GP is defined as  
\begin{align} 
\label{GP_general}
& \min_{\boldsymbol{\textbf{x}}}  f_{0}(\textbf{x}),   \\
& \text{subject to:}  \nonumber \\
& f_{i}(\textbf{x})\leq{1}, \quad \forall i= 0,1,\dots, I, \nonumber \\
& g_{j}(\textbf{x})={1}, \quad \forall j= 0,1,\dots, J,\nonumber
\end{align}
where $\textbf{x}=[x_{1}, x_{2},..., x_{N}]$ is a non-negative optimization variable vector, $g_{j}(\textbf{x})$ for all $j$ is a monomial function, i.e., $g_{j}(\textbf{x})=\prod\limits_{n=1}^Nc_{jn}x_{n}^{a_{jn}}$ where $c_{jn}>{0}$, $a_{jn}\in{\Re}$, and $f_{0}(\textbf{x})$ and $f_{i}(\textbf{x})$ for all $i$ are posynomial functions, i.e., $f_i(\textbf{x})=\sum_{k=1}^{K_i}\prod\limits_{n=1}^Nc_{ikn}x_{n}^{a_{ikn}}$. However, in \eqref{GP_general}, there are a lot of restrictions on the equality and inequality constraints which cannot be met for many practical problems related to the resource allocation of wireless networks such as the optimization problem considered in this paper. For example, in some cases, the equality constraints contain posynomial functions, inequality constraints present lower bound of posynomial function or the posynomial functions contain negative coefficients. Depending on the nature of the optimization problem, these types of problems belong to either one of classes of optimization problems such as generalized GP, signomial programming or complementary geometric programming (CGP). A CGP can be presented as 
\begin{align}\label{CGP_general}
& \min_{\boldsymbol{\textbf{x}}} F_{0}(\textbf{x}),   \\
& \text{subject to:}  \nonumber \\
& F_{i}(\textbf{x}) \leq{1}, \quad \forall i= 1,\dots, I,   \nonumber\\
& G_{j}(\textbf{x})={1}, \quad \forall j= 1,\dots, J,\nonumber
\end{align}
where $F_{0}(\textbf{x})=f_0^+(\textbf{x})-f_0^-(\textbf{x})$, $F_{i}(\textbf{x})=\frac{f_i^+(\textbf{x})}{f_i^-(\textbf{x})}$ for all $i=1, \cdots, I$ and $G_{j}(\textbf{x})=\frac{g_j(\textbf{x})}{f_j(\textbf{x})}$ in which $f_0^+(\textbf{x})$, $f_0^-(\textbf{x})$ for all $i=0,1, \cdots, I,$ are posynomial functions, and $g_j(\textbf{x})$ and $f_j(\textbf{x})$ are monomial and posynomial functions for all $j =1, \cdots J,$ respectively, \cite{avriel1970complementary}. 

Toward solving \eqref{CGP_general}, one approach is to convert the CGP optimization problem into a sequence of standard GP problems \cite{xu2014global} that can be
solved to reach a global solution. In other words, successive convex approximation (SCA) is applied \cite{SCA,SCA2}, where the non-convex optimization problem is approximated as a convex problem in each iteration. Lets consider $l$ as an iteration number for this sequence.  More specifically, arithmetic-geometric mean approximation (AGMA) can be applied to transform the non-posynomial functions to posynomial form, i.e., $F_{i}(\textbf{x})$, and $G_{j}(\textbf{x})$ to its monomial functions, respectively. 

Using AGMA, at the iteration $l$, the approximated forms of $f_i^-(\textbf{x})=\sum_{k=1}^{K^-_i}g^{i^-}_{k}(\textbf{x})$ and $f_j(\textbf{x})=\sum_{k=1}^{K_j}g^j_{k}(\textbf{x})$ are
\begin{align} 
\label{AGMA1} & \widetilde{f}^-_i(\textbf{x}(l))=\prod_{k=1}^{K_{i^-}}\left(\frac{g_{k}^{i^-}(\textbf{x}(l))}{\alpha_{k}^{i^-}(l)}\right)^{\alpha_k^{i^-}(l)},
\end{align} 
and
\begin{align}\label{AGMA2}  & \widetilde{f}_j(\textbf{x}(l))=\prod_{k=1}^{K_j}\left(\frac{g^j_{k}(\textbf{x}(l))}{\zeta_{k}^j(l)}\right)^{\zeta_k^j(l)},\
\end{align}
where $\alpha_k^{i^-}(l)=\frac{g_{k}^{i^-}(\textbf{x}(l-1))}{f_i^-(\textbf{x}(l-1))}$ and $\zeta_k^j(l)=\frac{g_{k}^j(\textbf{x}(l-1))}{f_j(\textbf{x}(l-1))}$. Now, we have $\widetilde{F}_{i}(\textbf{x}(l))=\frac{f_i^+(\textbf{x}(l))}{\widetilde{f}^-_i(\textbf{x}(l))(\textbf{x}(l))}$ and $\widetilde{G}_{j}(\textbf{x}(t))=\frac{g_j(\textbf{x}(t))}{\widetilde{f}_j(\textbf{x}(t))}$ which are posynomial and monomial functions, respectively \cite{xu2014global}. Now, the optimization problem related to each iteration $l$ of \eqref{CGP_general} is

\begin{align}\label{CGP_general2}
& \min_{\boldsymbol{\textbf{x}(l)}} \,\,\,\,\,\, \Xi +f_0^+(\textbf{x}(l))-f_0^-(\textbf{x}(l)),   \\
& \text{subject to:}  \nonumber \\
& \widetilde{F}_{i}(\textbf{x}(l)) \leq{1}, \quad \forall i= 1,\dots, I,   \nonumber\\
& \widetilde{G}_{j}(\textbf{x}(l))={1}, \quad \forall j= 1,\dots, J,\nonumber
\end{align}
where $\Xi >0$ is a sufficiently large constant which is added to the objective optimization problem \eqref{CGP_general2} to ensure that the objective function is always positive \cite{xu2014global}. However, the objective function of \eqref{CGP_general2} still cannot satisfy the posynomial condition of \eqref{GP_general}. To take the final step to reach the GP based formulation for each iteration, we introduce the auxiliary variable $x_0>0$ to reach a linear objective function and apply it for transformation of \eqref{CGP_general2} into 
\begin{align}\label{CGP_general-2}
& \min_{\textbf{x}_0(t)} x_{0}(l),   \\
& \text{subject to:}  \nonumber \\
&\frac{\Xi +f_0^+(\textbf{x}(l))}{f_0^-(\textbf{x}(l))+x_0}\leq 1,\nonumber\\
& \widetilde{F}_{i}(\textbf{x}(t)) \leq{1}, \quad \forall i=0, 1,\dots, I,  \nonumber\\
& \widetilde{G}_{j}(\textbf{x}(t))={1}, \quad \forall j= 1,\dots, J,\nonumber
\end{align}
where $\textbf{x}_0(t)=[x_{0}(l), x_{n}(l), \cdots, x_{0}(l)]$. Similar to $F_{i}(\textbf{x})$, term $\frac{\Xi +f_0^+(\textbf{x}(l))}{f_0^-(\textbf{x}(l))+x_0}$ can be converted into posynmial function via AGMA, and finally, the resulting optimization problem has a GP based structure and can be solved by efficient numerical algorithm or available softwares \cite{xu2014global}. 

It has been shown that the optimal solution of iterative algorithm based on the approximation of problem \eqref{CGP_general} into its GP based approximation has a very close performance to the optimal solution \cite{xu2014global}. 
\subsection{User Association Problem }
Considering a fixed $\textbf{P}(t)$, we have the following optimization problem, referred to as the user association optimization problem,
\begin{align} \label{User Association}
	& \max_{\boldsymbol{\beta}}\sum_{m\in \mathcal{M}}\sum_{g\in \mathcal{G}}\sum_{n_g\in\mathcal{N}_g}\sum_{k\in\mathcal{K}}\beta_{m,k,n_g}R_{m,k,n_g}(\textbf{P}(t)),   \\
	& \text{subject to:} \quad   \widetilde{\text{C1}}, \text{C3, C4,} \nonumber
\end{align}
where $R_{m,k,n_g}(\textbf{P}(t))$ has a fixed value derived based on $\textbf{P}(t)$ and 
\begin{equation*}\label{C1}
\begin{split}
\widetilde{\text{C1}}: \quad \quad & \sum_{m\in \mathcal{M}}\sum_{n_g\in\mathcal{N}_g}\sum_{k\in \mathcal{K}}{\beta_{m,k,n_g}}R_{m,k,n_g}(\textbf{P}(t))\geq{R_g^\text{rsv}}, \quad \\
& \forall g \in \mathcal{G}.
\end{split}
\end{equation*}

In \eqref{User Association}, the only optimization variable is $\boldsymbol{\beta}$, and therefore, \eqref{User Association} has less computational complexity compared to \eqref{original_problem}. However, it still suffers from the integer optimization variable $\boldsymbol{\beta}$. In addition, due to C4 and the objective function, \eqref{User Association} is still a non-convex optimization problem. To overcome the computational complexity of \eqref{User Association}, we first relax the UAF variable as $\beta_{m,k,n_g} \in [0,1]$. Also, we apply the technique in Section III. A to convert \eqref{User Association} into the GP formulation. 

\setlength{\parskip}{1mm plus0mm minus0mm}
To reach to a standard GP formulation, the equality constraint in C4 should involve only monomial functions. In the following, we first relax and then apply iterative AGMA algorithm (as in \eqref{AGMA1} and \eqref{AGMA2}) to get the monomial approximation for C4. Also, we show how we can convert the objective function of \eqref{AGMA1} into the standard form of GP. \vspace{-1 mm}

\textbf{Proposition 1}: Consider $t_1$ as the index of iteration for solving \eqref{User Association},  $x_{m,n_g}(t_1)=\sum_{k\in{\mathcal{K}}}\beta_{m,k,n_g}(t_1)$ and $y_{n_g}(t_1)=\sum_{m\in{\mathcal{M}}}\sum_{k\in{\mathcal{K}}}\beta_{m,k,n_g}(t_1)$. C4 can be approximated by the following constraints.
\begin{align}
\text{C4.1: } \quad \quad  & \nonumber s_{m,n_g}^{-1}(t_1)+x_{m,n_g}(t_1)y_{n_g}(t_1)s_{m,n_g}^{-1}(t_1) \leq{1}, \quad \\ \nonumber
& \forall n_g \in \mathcal{N}_g, \, g \in \mathcal{G}, \, \forall m \in \mathcal{M}, \nonumber \\ \nonumber
\text{C4.2:} \quad \quad & \big[\frac{1}{\lambda_{m,n_g}(t_1)}\big]^ {-\lambda_{m,n_g}(t_1)}s_{m,n_g}(t_1) \times \\ \nonumber
& \bigg[\frac{x^2_{m,n_g}(t_1)}{\alpha_{m,n_g}(t_1)}\bigg]^{-\alpha_{m,n_g}(t_1)} \leq 1, \quad \\
& \forall n_g \in \mathcal{N}_g, \, g \in \mathcal{G}, \, \forall m\in \mathcal{M},\nonumber \\  \nonumber
\text{C4.3: } \quad \quad & x_{m,n_g}(t_1)\prod_{k\in{K}}  \left[\frac{\beta_{m,k,n_g}(t_1)}{\nu_{m,k,n_g}(t_1)}\right]^{-\nu_{m,k,n_g}(t_1)}=1, \quad \\ \nonumber
& \forall n_g \in \mathcal{N}_g, \, g \in \mathcal{G}, \, \forall m \in \mathcal{M},\nonumber \\ \nonumber
\text{C4.4: } \quad \quad 
& y_{n_g}(t_1)\prod_{m\in{\mathcal{M}},k\in{\mathcal{K}}}\left[\frac{\beta_{m,k,n_g}(t_1)}{\eta_{m,k,n_g}(t_1)}\right]^{-\eta_{m,k,n_g}(t_1)} =1,  \quad \\ \nonumber
& \forall n_g \in \mathcal{N}_g, \, g \in \mathcal{G}, \, \forall m \in \mathcal{M},
\end{align} 
\vspace{10mm}
where $ s_{m, n_g}(t_1) $ is an auxiliary variable, and,
\vspace{-7mm}
\allowdisplaybreaks
\begin{align}
\label{13-31}
& \lambda_{m,n_g}(t_1) = \frac{1}{x^2_{m,n_g}(t_1 -1) + 1}, \\
& \alpha_{m,n_g}(t_1) = \frac{x^2_{m,n_g}(t_1 -1)}{x^2_{m,n_g}(t_1 -1) + 1}, \quad \\
\label{12} 
& \nu_{m,k,n_g}(t_1)=\frac{\beta_{m,k,n_g}(t_1-1)}{\sum_{k\in{\mathcal{K}}}\beta_{m,k,n_g}(t_1-1)}, \quad , \\ \label{13}
&\eta_{m,k,n_g}(t_1)=\frac{\beta_{m,k,n_g}(t_1-1)}{\sum_{m\in{\mathcal{M}}}\sum_{k\in{\mathcal{K}}}\beta_{m,k,n_g}(t_1-1)}, \quad \\ \nonumber
& \qquad \qquad \qquad \forall n_g \in \mathcal{N}_g, \, g \in \mathcal{G}, \, \forall m \in \mathcal{M}.
\end{align}
%
\begin{proof}
See Appendix A. 
\end{proof}
Based on C4.1-C4.4, C4 is transformed and represented by approximated monomial equalities and posynomial inequalities. Next, we show how we can transform the objective function into the monomial function to reach the GP based formulation for \eqref{User Association}. 

\textbf{Proposition 2}: Consider auxiliary variable $x_0>0$ and $\Xi_1 \gg 1$. The user association problem \eqref{User Association} at each iteration $t_1$ can be transformed into the following standard GP problem

\begin{align}
\label{beta_eqn}
& \min_{\boldsymbol{\beta}(t_1), \, x_0(t_1), \, \textbf{s}_{m,n_g}(t_1)} x_0(t_1),  \\
& \text{subject to}: \quad \text{C4.1-C4.4}, \nonumber \\
& \Xi_1 \left[\frac{x_0(t_1)}{c_0(t_1)}\right]^{-c_0(t_1)} \prod_{m\in\mathcal{M},g\in\mathcal{G},n_g\in{ \mathcal{N}_g},k\in \mathcal{K}} \\ \nonumber
& \left[\frac{\beta_{m,k,n_g}(t_1){R_{m,k,n_g}(\textbf{P}(t))}}{c_{m,k,n_g}(t_1)}\right]^{-c_{m,k,n_g}(t_1)}\leq 1, \nonumber \\ \nonumber
& \widetilde{\text{C1.1}}: \quad  R_{g}^{\text{rsv}} \times \\ \nonumber
& \prod_{m\in\mathcal{M},n_g\in{ \mathcal{N}_g},k\in \mathcal{K}}\left[\frac{\beta_{m,k,n_g}(t_1)R_{m,k,n_g}(\textbf{P}(t))}{\varphi_{m,k,n_g}(t_1)}\right]^{-\varphi_{m,k,n_g}(t_1)}\leq{1}, \quad\\
& \forall{g\in{\mathcal{G}}}, \nonumber \\
& \text{C3.1:}    \quad \sum_{g\in{\mathcal{G}}}\sum_{n_{g}\in{\mathcal{N_{G}}}}\beta_{m,k,n_g}(t_1)\leq1, \quad \forall{m} \in \mathcal{M}, \forall{k} \in \mathcal{K}, \nonumber 
\end{align}
where
\begin{align}
\label{user_assoc_eqn} &\varphi_{m,k,n_g}(t_1)= \\ \nonumber
& \frac{\beta_{m,k,n_g}(t_1-1)R_{m,k,n_g}(\textbf{P}(t))}{\sum_{m\in{\mathcal{M}}}\sum_{n_g\in{\mathcal{N}_g}}\sum_{k\in{\mathcal{K}}}\beta_{m,k,n_g}(t_1-1)R_{m,k,n_g}(\textbf{P}(t))},\quad  \\ \nonumber
& \forall{g\in{\mathcal{G}}}, \end{align}
and $ c_{m, k, n_g}(t_1) $ and $ c_{0}(t_1) $ are defined in \eqref{7} and \eqref{C0} at the top of the next page.

\begin{proof}
See Appendix B. 
\end{proof}

Now, at each iteration, the optimization problem can be replaced by its GP approximation in \eqref{beta_eqn}. Iteratively, \eqref{beta_eqn} will be solved till the optimal value of UAF will be obtained. The overall iterative algorithm to solve \eqref{beta_eqn} is summarized in Algorithm 1.A where the iterative algorithm is repeated till the convergence condition holds. 

\begin{algorithm}[t]
	\renewcommand{\thealgorithm}{}
	\caption{\textbf{1.A }: Algorithm to derive User Association Factor}
	\label{Algorithm_2}
		 \textbf{Initialization:}
		 Set $t_1 :=1$, $\boldsymbol{\beta}(t_1=1) =\boldsymbol{\beta}(t)$ and $\textbf{P}(t_1=1)=\textbf{P}(t)$ and set arbitrary initial for $s_{m,n_g}(t_1)$, \\
		\textbf{Repeat}  
		
		 \hspace{.5cm}\textbf{Step 1: }Update $ \gamma_{m,n_g}(t_1), \alpha_{m,n_g}(t_1), \nu_{m,k,n_g}(t_1), \\ \eta_{m,k,n_g}(t_1), c_{m,k,n_g}(t_1), c_0(t_1)$ and $\varphi_{m,k,n_g}(t_1)$ from \eqref{13-31}-\eqref{13} and \eqref{user_assoc_eqn}-\eqref{C0},\\

		 \hspace{.5cm}\textbf{Step 2: }Find optimal UAF in \eqref{beta_eqn} using CVX,\\
		 
		 \hspace{.5cm}\textbf{Step 3: } Stop if $\vert\vert{\boldsymbol{\beta}}^*(t_1)-\boldsymbol{\beta}^*(t_1-1)\vert\vert\leq{\varepsilon}_1$, else set $t_1:=t_1+1$ and go to \textbf{Step 1}.
\end{algorithm}

\begin{figure*}[t] 
	\begin{minipage}[t]{\linewidth}
		\begin{@twocolumnfalse}
			\begin{align}\label{7} 
			& c_{m,k,n_g}(t_1)= \frac{{\beta_{m,k,n_g}(t_1-1){R_{m,k,n_g}(\textbf{P}(t))}}}{x_0(t_1-1)+\sum_{m\in \mathcal{M}}\sum_{g\in\mathcal{G}}\sum_{n_g\in\mathcal{N}_g}\sum_{k\in \mathcal{K}}{\beta_{m,k,n_g}(t_1-1){R_{m,k,n_g}(\textbf{P}(t))}}}, \\ 
			& c_0(t_1)=\frac{x_0(t_1-1)}{x_0(t_1-1)+\sum_{m\in \mathcal{M}}\sum_{g\in\mathcal{G}}\sum_{n_g\in\mathcal{N}_g}\sum_{k\in \mathcal{K}}{\beta_{m,k,n_g}(t_1-1){R_{m,k,n_g}(\textbf{P}(t))}}}. \label{C0}
			\end{align}
			\vspace{-5mm}
			\begin{center}
				\line(1,0){515}
			\end{center}
		\end{@twocolumnfalse}
	\end{minipage}
\end{figure*}

\textbf{Proposition 3}: With AGMA, Algorithm 1.A converges to a locally optimal solution that satisfies the KKT conditions of the original problem.
\begin{proof}
	In \cite{4275017}, it is shown that the conditions for the 	convergence of the SCA \cite{SCA} are satisfied and guarantee that the solutions of the series of approximations by AGMA converges to a point that satisfies the KKT
	conditions of \ref{User Association}, i.e., a local maximum is attained.
\end{proof}

\subsection{Power Allocation Problem}
For a given set of UAFs derived from Algorithm A.1., the optimization problem of power allocation becomes 
\begin{align}
\label{power_rate_eqn}
& \max_{\textbf{P}(t_2)}\sum_{m\in \mathcal{M}}\sum_{g\in\mathcal{G}}\sum_{n_g\in\mathcal{N}_g}\sum_{k\in \mathcal{K}}\beta_{m,k,n_g}(t)R_{m,k,n_g}(\textbf{P}(t_2))\\
& \text{subject to:} \nonumber \\
\widetilde{\text{C1.2}}: & \sum_{m\in{\mathcal{M}}}\sum_{k\in{\mathcal{K}}}\sum_{n_g\in{\mathcal{N}_g}}\beta_{m,k,n_g}(t)R_{m,k,n_g}(\textbf{P}(t_2))\geq{R_g^\text{rsv}}, \\ \nonumber
& \forall{g}\in{\mathcal{G}}, \nonumber \\ \nonumber
\widetilde{\text{C2.2}}:& \sum_{g\in{\mathcal{G}}}\sum_{n_{g}\in{\mathcal{N}_g}}\sum_{k\in{\mathcal{K}}}P_{m,k,n_g}(t_2)\leq{P_{m}^{\text{max}}},\quad \forall{m}\in{\mathcal{M}}, \nonumber
\end{align}
where $t_2$ is the index of iterations. Due to interference in the objective function of $R_{m,k,n_g}(\textbf{P}(t_2))$, \eqref{power_rate_eqn} is a non-convex optimization problem. We again follow the approach of Section III.A to convert \eqref{power_rate_eqn} into the GP optimization problem. First, we rewrite the objective of \eqref{power_rate_eqn} as
\begin{align}
\label{cons_3}
\prod_{m\in\mathcal{M},g\in\mathcal{G},n_g\in{ \mathcal{N}_g},k\in \mathcal{K}}\gamma_{m,k,n_g}(\textbf{P}(t_2)) 
\end{align}
where $$\gamma_{m,k,n_g}(\textbf{P}(t_2))=\frac{\sigma^2+I_{m,k,n_g}(t_2)+P_{m,k,n_g}(t_2)h_{m,k,n_g}}{\sigma^2+I_{m,k,n_g}(t_2)}$$ and 
\begin{align} \nonumber
& I_{m,k,n_g}(t_2)= \\ \nonumber
& \sum_{\forall{m'}\in\mathcal{M}, m'\neq{m}}\sum_{\forall{g}\in\mathcal{G}}\sum_{\forall{n_g'}\in\mathcal{N}_g, n_g'\neq{n_g}}P_{m',k,n'_g}(t_2)h_{m, k,n'_{g}}.
\end{align}

Now from AGMA in Section III.B, $\gamma_{m,k,n_g}(\textbf{P}(t_2))$ can be approximated as
\begin{align*}
& \widehat{\gamma}_{m,k,n_g}(\textbf{P}(t_2)) = (\sigma^{2}+I_{m,k,n_g}(t_2))\left(\frac{\sigma^{2}}{\kappa_{o}(t_2)}\right)^{-\kappa_{o}(t_2)} \\
& \prod_{m\in\mathcal{M},g\in\mathcal{G},n_g\in{ \mathcal{N}_g},k\in \mathcal{K}}\left({\frac{P_{m,k,n_g}(t_2)h_{m,k,n_g}}{\kappa_{m,k,n_g}(t_2)}}\right)^{-\kappa_{m,k,n_g}(t_2)},
\end{align*}
where  
\begin{align*}
& \kappa_{m,k,n_g}(t_2)= \\ 
& \frac{P_{m,k,n_g}(t_2-1)h_{m,k,n_g}}{\sigma^{2}+\sum_{m\in{\mathcal{M}},n_g\in{\mathcal{N}_g},g\in{\mathcal{G}}}P_{m,k,n_g}(t_2-1)h_{m,k,n_g}},
\end{align*}   and 
\begin{align*}
\kappa_o(t_2)=\frac{\sigma^{2}}{\sigma^{2}+\sum_{m\in{\mathcal{M}},n_g\in{\mathcal{N}_g},g\in{\mathcal{G}}}P_{m,k,n_g}(t_2-1)h_{m,k,n_g}}.
\end{align*}
Consequently, \eqref{power_rate_eqn} is transformed into the following standard GP problem   
\begin{align}
	\label{power_eqn}
	& \min_{\textbf{P}(t_2)} \prod_{m\in\mathcal{M},g\in\mathcal{G},n_g\in{ \mathcal{N}_g},k\in \mathcal{K}}\widehat{\gamma}_{m,k,n_g}(\textbf{P}(t_2))\\
	& \text{subject to:} \nonumber \\ \nonumber	
	\widetilde{\text{C1.2}}:& \prod_{m\in\mathcal{M},g\in\mathcal{G},n_g\in{ \mathcal{N}_g},k\in \mathcal{K}}\widehat{\gamma}_{m,k,n_g}(\textbf{P}(t_2))\leq{2^{-R_{g}^{\text{rsv}}}},  \\ \nonumber
	& \qquad \forall{g}\in{\mathcal{G}}, \nonumber \\
	\widetilde{\text{C2.2}}: & \sum_{g\in{\mathcal{G}}} \sum_{n_{g}\in{\mathcal{N}_{g}}}\sum_{k\in{\mathcal{K}}}P_{m,k,n_g}(t_2) \leq P_{m}^{\text{max}},\quad \forall{m}\in{\mathcal{M}}, \nonumber \nonumber
\end{align} 
The overall optimization problem is iteratively solved as described in Algorithm 1.B until the power vector converges, i.e., $\vert\vert\textbf{P}(t_2)-\textbf{P}(t_2-1)\vert\vert\leq{\varepsilon_2}$ where $0< \varepsilon_2 \ll 1$. Note that Proposition III holds for the  iterative Algorithm 1.B.

\begin{algorithm}[t]
	\renewcommand{\thealgorithm}{}
	\caption{\textbf{1.B:} Algorithm to Derive Power Allocation }
	\label{Algorithm_3}
	 \textbf{Initialization:} Set $t_2:=1$ and obtain $\boldsymbol{\beta}(t_2=1)=\boldsymbol{\beta}(t)$ from Algorithm \textbf{1.A}.\\
	 \textbf{Repeat} \\		 
\textbf{Step 1:} Update $\kappa_{m,k,n_g}(t_2)$ and $\kappa_o(t_2)$, \\	
		\textbf{Step 2: }Find optimum power allocation according to \eqref{power_eqn} using CVX \cite{cvx,boyd2009convex},\\
		 \textbf{Step 3:} Stop if $\vert\vert\textbf{P}(t_2)-\textbf{P}(t_2-1)\vert\vert\ll{\varepsilon_2}$, else set $t_2:=t_2+1$ and go to \textbf{Step 1}.
\end{algorithm}

\subsection{Computational Complexity}

When CVX is applied for the problems in Algorithms 1.A and 1.B, it solves GP problems with the interior point method. Given by \cite{boyd2009convex}, the number of required iterations is equal to 
$\frac{\log(c /(t^0\varrho))}{\log(\xi)}$, 
where $c$ is the total number of constraints in \eqref{beta_eqn}, $t^0$ is the initial point for approximating the accuracy of interior point method, $0<\varrho\ll 1$ is the stopping criterion for interior point method, and $\xi$ is used for updating the accuracy of interior point method \cite{boyd2009convex}. In the problem described earlier, the number of constraints in \eqref{beta_eqn} is $c_1 = G + MK + 4MN+1$ for Algorithm 1.A and $c_2= G + M$
for Algorithm 1.B.

Moreover, in Algorithms 1.A and 1.B, for each iteration, the number of computations required to convert the non-convex problems using AGMA into \eqref{beta_eqn} and \eqref{cons_3} is $i_1 = K M^2 N+6 K M N+MKG N$ and $i_2 =GMKN+2MKN$, respectively. Therefore, the total number of computations for Algorithms 1.A and 1.B are 
 \begin{equation*}
 \text{Computational complexity} = \begin{cases}
 i_1 \times \frac{\log(c_1 /(t_1^0\varrho_1))}{\log(\xi_1)}, \\
 \text{ for Algorithm 1. A},\\
 i_2 \times \frac{\log(c_2 /(t_2^0\varrho_2))}{\log(\xi_2)}, \\
 \text{ for Algorithm 1. B.}
 \end{cases}
 \end{equation*}
  
Based on this analysis, the computation complexity of Algorithm 1.A is significantly higher than that of Algorithm 1.B. Moreover, Algorithm 1.A is more sensitive to $K$ and $N$ than Algorithm 1.B. 
\section{Simulation results}
\subsection{Simulation Parameters}
We consider a multi-cell VWN scenario with $M=4$, $K=4$ and $G=2$. The radius of each BS is considered to be $500$ m where the users are randomly located (from a uniform distribution) within the whole area of interest unless otherwise stated. The channel gains are derived by the Rayleigh fading model in which $h_{m,n_g,k}=\chi_{m,n_g,k} d_{m,n_g}^{-\beta}$ where $\beta=3$ is the path loss exponent, $d_{m,n_g}$ is the distance between the BS $m$ and user $n_g$ and $\chi_{m,n_g,k}$ is the exponential random variable with mean/parameter equal to $1$. We set $N_g=4$ and $P_m^{\text{max}}=40$ dBm, unless otherwise stated. Also, for all of the simulations, we set $\Xi_1 = 10^{7} $ and $\varepsilon_1 =10^{-5}$, $\varepsilon_2 = 10^{-6}$. In all of the following simulations, for each realization of network, when there exists no feasible solution for the system, i.e., C1-C4 cannot be hold simultaneously, the corresponding total rate is set to be zero. Also for all the following simulations, we have $R^{\text{rsv}}=R_{g}^{\text{rsv}}$ for all $g \in \mathcal{G}$ and $P^{\text{max}}=P_{m}^{\text{max}}$ for all $m \in \mathcal{M}$. To compare the performance of our algorithm, we apply the traditional algorithm explained in the following subsection.
 
\subsection{Traditional Approach}
In practical scenario of wireless networks, e.g., 2G, the users are assigned to BSs based on the highest average received SINR measured during one measurement window. 
In this case, the resource allocation problem is transformed into 
\begin{align} \label{original_problem2}
& \max_{\boldsymbol{\beta}',\textbf{P}}\sum_{m\in \mathcal{M}}\sum_{g\in \mathcal{G}}\sum_{n_g\in\mathcal{N}_g}\sum_{k\in\mathcal{K}}\beta_{m,k,n_g}'R_{m,k,n_g}  (\textbf{P}) \\
& \text{subject to: C1 - C3},  \nonumber
\end{align}
where $\boldsymbol{\beta}'=[\beta_{m,k,n_g}']_{\forall m, k, n_g}$ and $\beta_{m,k,n_g}'$ shows the sub-carrier allocation of user $n_g$ on sub-carrier $k$ when it is allocated to the BS $m$. To solve \eqref{original_problem2}, we can apply the algorithm similar to Algorithm 1 introduced in Section III, except that here, \eqref{beta_eqn} contains C1-C3, and C4 are removed. When the sub-carrier assignment is solved, the optimal power is derived from Algorithm 1.B, for \eqref{original_problem2}. This iterative algorithm is terminated when the convergence conditions are met which is summarized in Algorithm 2.

\begin{figure}
	\centering
	\mbox{\subfloat[]{\label{Fig2}
			\includegraphics[width=0.9\linewidth]{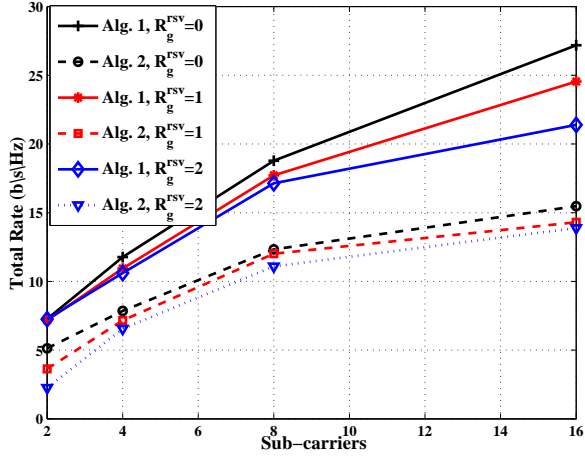}}}	
	\mbox{\subfloat[]{\label{Fig3} \includegraphics[width=\linewidth]{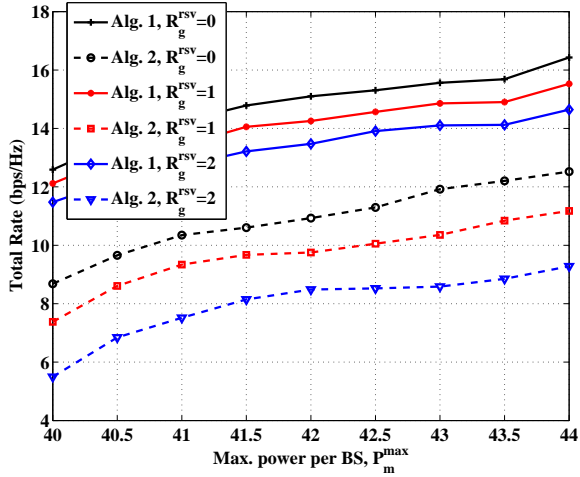}}}
	\caption{Total rate versus (a) sub-carriers, $K$, and (b) maximum transmit power per BS, $P^{\text{max}}$}
\end{figure}

\begin{figure}[!hbt]
	\centering
	\mbox{\subfloat[]{\label{OP}
			\includegraphics[scale = 0.5]{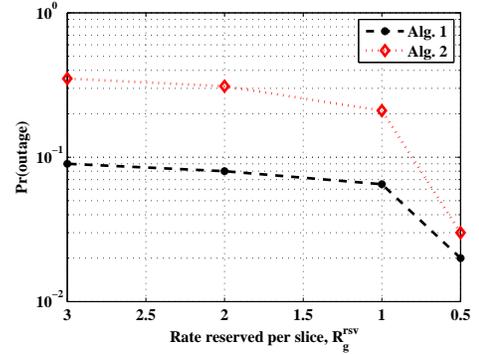}}}	
	\mbox{\subfloat[]{\label{Fig4} \includegraphics[scale = 0.5]{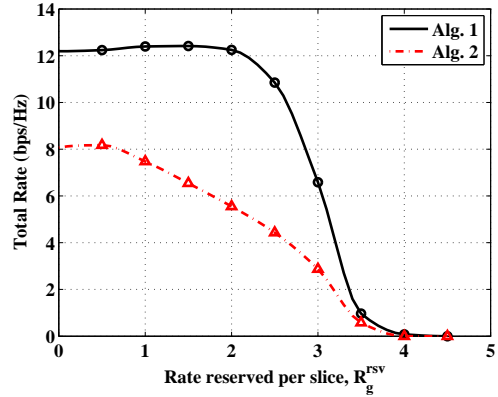}}}
	\caption{$R^{\text{rsv}}$ versus (a) outage probability (Pr(outage)), and (b) total rate}
\end{figure}

\begin{algorithm}
	\renewcommand{\thealgorithm}{}
	\caption{\textbf{2 }}
	\label{Algorithm_4}
		 \textbf{Initialization:} Set $t_{3} :=1$,  BS assignment where user $n_g$ is assigned to BS $m$ based on the average received SINR. \\
		 		 \textbf{Repeat}\\
		 		 \hspace{.5cm}\textbf{Step 1: } Compute $\boldsymbol{\beta}^{'*}(t_{3})$ from Algorithm 1.B except that BS is assigned based on the signal strength.\\
		 		 \hspace{.5cm}\textbf{Step 2: } For a fixed $\boldsymbol{\beta}^{'*}(t_{3})$, find the optimal power allocation $\textbf{P}(t_3)$ from Algorithm \textbf{1.B}.\\
		 		 \hspace{.5cm}\textbf{Step 3: } Stop if $\vert\vert{\boldsymbol{\beta}}^{'*}(t_{3})-\boldsymbol{\beta}^{'*}(t_{3}-1)\vert\vert\leq{\varepsilon_1}  \text{ and }	 \vert\vert{\textbf{P}}^*(t_{3})-\textbf{P}^*(t_{3}-1)\vert\vert \leq \varepsilon_2$, where $0 <\varepsilon_2 \ll 1$, otherwise $t_3:=t_3+1$ and go to \textbf{Step 1}.
\end{algorithm}

\subsection{Evaluation of Algorithm 1 and Algorithm 2}

Primarily, we evaluate and compare the performances
of Algorithm 1 and Algorithm 2 versus different number of sub-carriers and maximum transmit power in Figs. \ref{Fig2} and \ref{Fig3}, respectively. The first point to notice from Figs. \ref{Fig2} and \ref{Fig3} is that Algorithm 1 considerably outperforms Algorithm 2 for different values of $R_{g}^{\text{rsv}}$, $K$ and $P_{m}^{\text{max}}$.

From Fig. \ref{Fig2}, it can be observed that the total rate is increased by increasing the total number of sub-carriers, i.e., $K$, due to the opportunistic nature of fading channels in wireless networks \cite{goldsmithbook}. As also expected, with increasing $P_{m}^{\text{max}}$, the total achievable rate is also increased which is shown in Fig. \ref{Fig3}. Both figures indicate that by increasing the value of $R_{g}^{\text{rsv}}$, the total rate decreases. It is because the feasibility region of resource allocation in \eqref{original_problem} is reduced by increasing $R_{g}^{\text{rsv}}$, leading to less total achieved throughput \cite{boyd2009convex}. However, from Fig. \ref{Fig3}, increasing $R_{g}^{\text{rsv}}$ has considerable effect on the performance of Algorithm 2 compared to Algorithm 1. It can be interpreted as Algorithm 1 can efficiently control interference between different cells compared to Algorithm 2. Therefore, the chance of feasible power allocation for larger values of $R_{g}^{\text{rsv}}$ is increased by Algorithm 1.

To study this point, we consider the outage probability of C1, expressed as 
\begin{align*}\label{C1}
\text{Pr(outage)}= & \text{Pr}\{\sum_{m\in \mathcal{M}}\sum_{n_g\in\mathcal{N}_g}\sum_{k\in\mathcal{K}}{\beta_{m,k,n_g}}R_{m,k,n_g}(\textbf{P},\boldsymbol{\beta}) \\
& \leq{R_g^\text{rsv}}\}, \quad\forall g \in \mathcal{G}. 
\end{align*}
Via Mont Carlo simulation, we derived Pr(outage) for the presented setup for both Algorithm 1 and Algorithm 2, depicted in Fig. \ref{OP}. In this figure, we assume $K=8$ and $P_{m}^{\text{max}}=40 \text{ dBm}$ for all $m \in \mathcal{M}$. The results demonstrate that as the rate reservation per slice $R_{g}^{\text{rsv}}$ increases, the outage probability for both Algorithm 1 and Algorithm 2 increases. However, Algorithm 2 has larger value of outage probability compared to the outage probability of Algorithm 1, meaning that the feasibility region of Algorithm 2 is smaller than that for Algorithm 1. On the other hand, Algorithm 1 can efficiently manage interference in the specific region between different cells compared to Algorithm 2. It is mainly because Algorithm 1 has more degree of freedom to choose the BS and allocate the sub-carriers among users of different slices. However, the BS assignment is predetermined in Algorithm 2. Therefore, the achieved rate derived by Algorithm 2 is less than that of {Algorithm 1}. Consequently, with increasing $R_{g}^{\text{rsv}}$, the decrease rate of Algorithm 2 is greater than that of Algorithm 1, since Algorithm 2 cannot manage the interference between different BSs. Hence, Algorithm 2 cannot satisfy the minimum required rate of slices leading to decrement of VWN efficiency.

For the same setup, in Fig. \ref{Fig4}, the total throughput of Algorithms 1 and 2 are plotted for different values of $R_{g}^{\text{rsv}}$. Fig. \ref{Fig4} clearly shows that Algorithm 1 yields higher throughput compared to Algorithm 2, while both of these algorithms become infeasible after $R_{g}^{\text{rsv}}=4$ bps/Hz. We should emphasize that in all simulation results, when the problem is infeasible, i.e., there is no power and sub-carrier vectors that can meet the constraint C1 for all $g \in \mathcal{G}$, the achieved total rate is set to zero. 
\begin{figure} 
	\centering
	\includegraphics[width=0.9\linewidth]{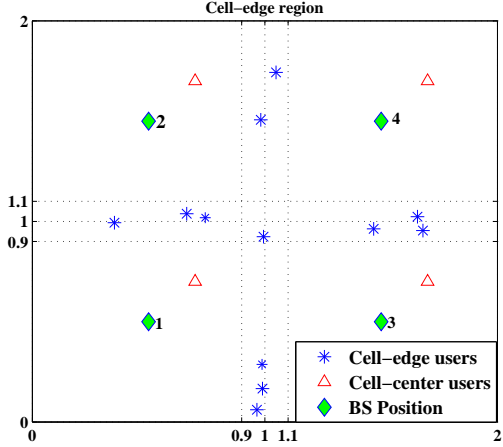} 	
	\caption{Illustration of network setup to investigate the coverage of multi-cell VWN}	\label{user_distribution}
\end{figure}

\subsection{Coverage Analysis}
In any cellular network, the coverage is one of the most important planning parameters which can be measured by SINR or achieved total rate of users at the cell boundaries. To study the performance of Algorithm 1 to increase the coverage, we consider the simulation setup similar to Fig. \ref{user_distribution} where the majority of users are located in the cell-edge, consequently, these users experience high interference from other BSs. Therefore, the achieved rate of each user is decreased which can be considered as the worst-case scenario of coverage analysis.

\begin{figure}
	\centering
	\subfloat[]{\label{CDF}
		\includegraphics[height=5.2cm]{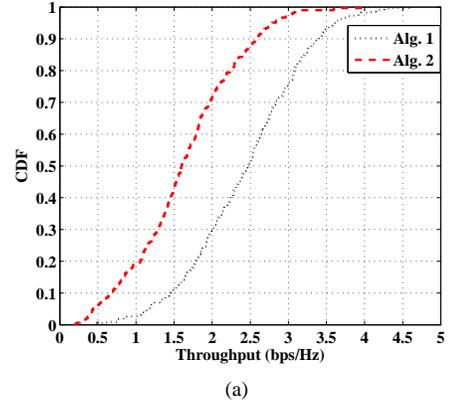}}\\
	\subfloat[]{\label{cell_center_CDF} \includegraphics[height=5.2cm]{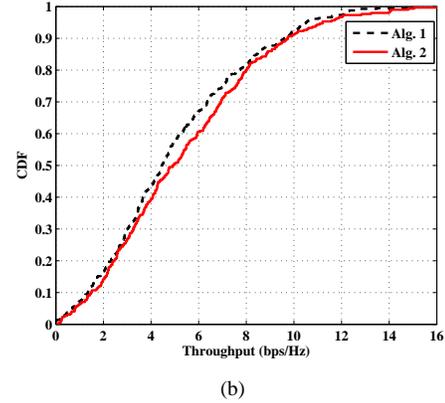}}
	\caption{Throughput distribution for (a) cell-edge, and (b) cell-center users}
\end{figure}
The cumulative distribution function (CDF) of the total throughput of cell-edge users and cell-center users are depicted in Figs. \ref{CDF} and \ref{cell_center_CDF}, respectively for both Algorithms 1 and 2. It can be seen that Algorithm 1 outperforms Algorithm 2 for the cell-edge users where 50\% of users in the cell-edge achieve a throughput of $2.5$ bps/Hz by Algorithm 1, while their throughput is around $1.5$ bps/Hz via Algorithm 2. However, the performance of both algorithms are similar for the cell-center users. It is because via user-association in Algorithm 1, the interference among different {cells} can be controlled while Algorithm 2 cannot control the interference through the connectivity of users to different BS and it is per-determined by the received SINR of reference signal. In other words, Algorithm 1 can provide better coverage even for cell-edge users for multi-cellular VWN which is desirable from implementation perspective.

\begin{figure}
	\centering
	\mbox{\subfloat[]{\label{uniform_users}
	\includegraphics[height=5cm]{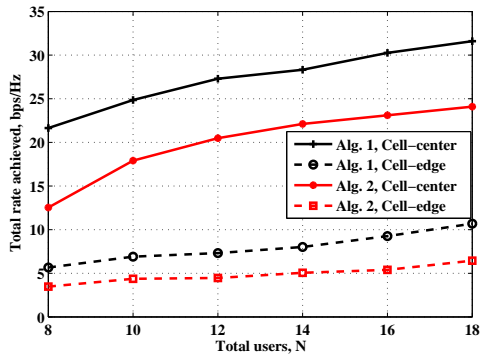}}}	
	\mbox{\subfloat[]{\label{nonuniform_users}  		\includegraphics[height=5cm]{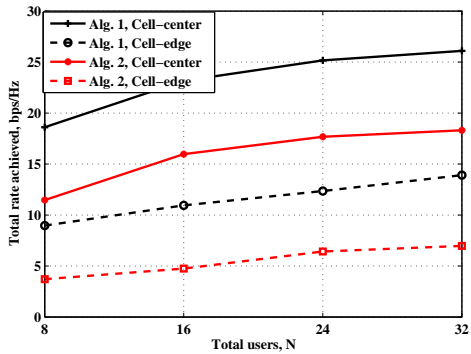}}}
	\caption{Total rate for (a) uniform user distribution, and (b) non-uniform user distribution}
\end{figure}

The performance is further investigated with respect to the number of users in the cell-edge in Figs. \ref{uniform_users} and \ref{nonuniform_users}. Obviously, Algorithm 1 can consistently improve performance of cell-edge users and at the same time, maintain desirable throughput of each slice regardless of the user deployment density compared to the Algorithm 2. For instance, for $N=18$, for the uniform user distribution, the total rate is increased from $7$ to $11$ bps/Hz and from $24$ to $32$ bps/Hz for cell-edge and cell-center users from Algorithm 2 to Algorithm 1, i.e., $57\%$ and $33\%$, respectively. For the non-uniform case, when $N=32$, the rate is incremented from $7$ to $12$ bps/Hz and $18$ to $27$ bps/Hz for cell-edge and cell-center users, respectively, which is equal to $71\%$ and $50\%$ increment in total rate from Algorithm 2 to Algorithm 1, respectively. These results {show} the efficiency of applying Algorithm 1 in increasing the coverage over the whole network. 

\subsection{Convergence Analysis of Algorithm 1}

In Fig. \ref{itn_a}, the number of iterations required for convergence for Algorithms 1.A and 1.B versus the total number of sub-carriers, $K$, is plotted for $N=8$ and $R_{g}^{\text{rsv}}=2$ bps/Hz. Similarly, in Fig. \ref{itn_b}, the number of iterations required for convergence versus the total number of users, $ N $ for $ K = 4 $ sub-carriers is plotted in the case of Algorithm 1.A and 1.B. According to the computational complexity analysis in Section III.D, as $N$ and $K$ increase, the number of iterations required for convergence also increases. The computational complexity in the case of Algorithm 1. A is more than that of Algorithm 1. B owing to the fact that the total number of constraints for Algorithm 1.A is higher than Algorithm 1.B which is also evident from the plots of iterations versus $N$ and $K$ in Fig. \ref{itn_a} and Fig. \ref{itn_b}.

\section{Conclusion}
In this paper, we proposed the joint BS, sub-carrier and power allocation algorithm for multi-cell OFDMA based virtualized wireless networks (VWNs). In the proposed setup, we have considered a set of slices (service providers), each has a set of their own users and require a minimum reserved rate. We formulated the related optimization problem based on the new defined optimization variable, called user association factor (UAF) indicating the joint sub-carrier and BS assignment. To solve the proposed non-convex and NP-hard optimization problem, we followed an iterative approach where in each step, one set of optimization variables is derived. However, even for each step, the optimization problem is non-convex and NP-hard. To derive the efficient algorithm to solve them, we apply the framework of iterative successive convex approximation via complementary geometric programming (CGP) to transform the non-convex optimization problem into the convex one. Then, to efficiently derive the solution, we applied CVX to solve the optimization problem of each step in this called two-layer iterative approach. Our simulation results reveal that via the proposed approach, the throughput and coverage of VWN, specifically for the cell-edge users, are considerably improved compared to the traditional scenario where the BS is assigned based on the maximum value of SINR.

\appendix
\subsection{Proof of Proposition 1}
From the definition of $x_{m,n_g}(t_1)$ and $y_{n_g}(t_1)$, C4 can be rewritten as 
\begin{equation}\label{transformC4}
\begin{split}
& x_{m,n_g}(t_1)[y_{n_g}(t_1)-x_{m,n_g}(t_1)]=0, \\
& \quad\forall{n_g\in \mathcal{N}_g},\,\,\forall{g}\in \mathcal{G},\,\, \forall{m} \in \mathcal{M}, 
\end{split}
\end{equation}
which is not a monomial function. \eqref{transformC4} can be rewritten as $x_{m,n_g}(t_1)y_{n_g}(t_1)=x^2_{m,n_g}(t_1)$ and by adding $1$ to both the left and right hand sides, we have $x_{m,n_g}(t_1)y_{n_g}(t_1)+1=1+x^{2}_{m,n_g}(t_1)$ for all ${n_g\in \mathcal{N}_g}$, ${g}\in \mathcal{G}$, and ${m} \in \mathcal{M}$. We define $s_{m,n_g}(t_1)\geq{0}$ as an auxiliary variable to relax and convert \eqref{transformC4} into the posynomial inequalities as follows \cite{GPmungchiang}
\begin{align}\label{transformC42} 
& x_{m,n_g}(t_1)y_{n_g}(t_1)+1\leq{s_{m,n_g}(t_1)}\leq{1+x^{2}_{m,n_g}(t_1)}, \quad \\ \nonumber
& \forall{n_g\in \mathcal{N}_g},\,\,\forall{g}\in \mathcal{G},\,\, \forall{m} \in \mathcal{M}.
\end{align}
The above inequalities can be written as 
\begin{align*}
& \frac{x_{m,n_g}(t_1)y_{n_g}(t_1)+1}{s_{m,n_g}(t_1)} \leq 1, \end{align*} and  \begin{align*}
& \frac{s_{m,n_g}(t_1)}{1+x^{2}_{m,n_g}(t_1)} \leq 1. 
\end{align*} 
\vspace{10mm}
Now, the above constraints can be approximated using AGMA approximation introduced in Section III. B. as 
\begin{align*}
\text{C4.1: } \quad & s_{m,n_g}^{-1}(t_1)+x_{m,n_g}(t_1)y_{n_g}(t_1)s_{m,n_g}^{-1}(t_1) \leq{1},\\
\text{C4.2:} \quad & \big[\frac{1}{\lambda_{m,n_g}(t_1)}\big]^ {-\lambda_{m,n_g}(t_1)} \times \\
& s_{m,n_g}(t_1)\bigg[\frac{x^2_{m,n_g}(t_1)}{\alpha_{m,n_g}(t_1)}\bigg]^{-\alpha_{m,n_g}(t_1)} \leq 1, 
\end{align*}
where $\lambda_{m,n_g}(t_1) = \frac{1}{x^2_{m,n_g}(t_1 -1) + 1}$ and $\alpha_{m,n_g}(t_1) = \frac{x^2_{m,n_g}(t_1 -1)}{x^2_{m,n_g}(t_1 -1) + 1}$. 
Now, C4 can be replaced by the following constraints 
\begin{align*}
& \text{C4.1: } \quad \quad  s_{m,n_g}^{-1}(t_1)+x_{m,n_g}(t_1)y_{n_g}(t_1)s_{m,n_g}^{-1}(t_1) \leq{1}, \\
& \text{C4.2:} \quad \quad  \big[\frac{1}{\lambda_{m,n_g}(t_1)}\big]^ {-\lambda_{m,n_g}(t_1)}s_{m,n_g}(t_1) \\
& \qquad \qquad\bigg[\frac{x^2_{m,n_g}(t_1)}{\alpha_{m,n_g}(t_1)}\bigg]^{-\alpha_{m,n_g}(t_1)} \leq 1, \\
& \widehat{\text{C4.3}}: \quad \quad x_{m,n_g}(t_1)=\sum_{k\in{\mathcal{K}}}\beta_{m,k,n_g}(t_1), \\
&  \widehat{\text{C4.4}}: \quad \quad y_{n_g}(t_1)=\sum_{m\in{\mathcal{M}},k\in{\mathcal{K}}}\beta_{m,k,n_g}(t_1),
\end{align*}
Note that via \eqref{transformC42}, the positive condition for the constraints of GP is met \cite{avriel1970complementary}. However, the equality constraints in $\widehat{\text{C4.3}}$ and $\widehat{\text{C4.4}}$ are not monomial since we have $x_{m,n_g}(t_1)-\sum_{k\in{\mathcal{K}}}\beta_{m,k,n_g}(t_1)=0$ and $y_{n_g}(t_1)-\sum_{m\in{\mathcal{M}},k\in{\mathcal{K}}}\beta_{m,k,n_g}(t_1)=0$, and, they have negative constraints. To convert $\widehat{\text{C4.3}}$ and $\widehat{\text{C4.4}}$ to the monomial functions, we again apply AGMA approximation presented in Section III.A as 
\begin{align}
\label{cons_1}
\text{C4.3: } \quad x_{m,n_g}(t_1)\prod_{k\in{K}}  \left[\frac{\beta_{m,k,n_g}(t_1)}{\nu_{m,k,n_g}(t_1)}\right]^{-\nu_{m,k,n_g}(t_1)}=1  
\end{align}
\begin{align}
\label{cons_2}
\text{C4.4: } \quad y_{n_g}(t_1)\prod_{m\in{\mathcal{M}},k\in{\mathcal{K}}}\left[\frac{\beta_{m,k,n_g}(t_1)}{\eta_{m,k,n_g}(t_1)}\right]^{-\eta_{m,k,n_g}(t_1)} =1,
\end{align}
where $\nu_{m,k,n_g}(t_1)$ and 
$\eta_{m,k,n_g}(t_1)$ are defined in \eqref{12} and \eqref{13}, respectively. 

\begin{figure}
	\centering
	\mbox{\subfloat[]{\label{itn_a}
			\includegraphics[width=0.8\linewidth]{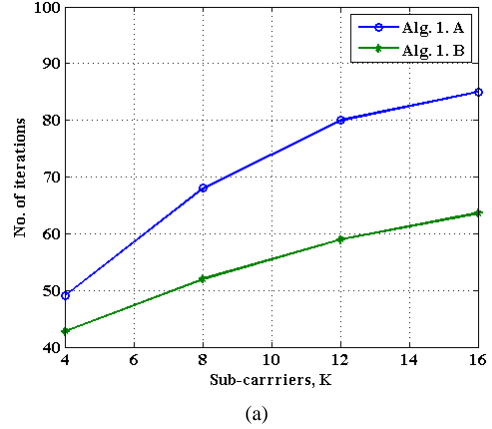}}}	
	\mbox{\subfloat[]{\label{itn_b} \includegraphics[width=0.8\linewidth]{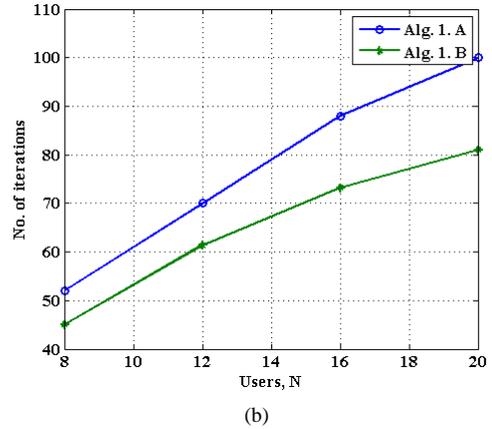}}}
	\caption{Number of required iterations for lower-level iterative algorithms versus (a) sub-carriers, $K$, and (b) total number of users, $N$}
\end{figure}

\subsection{Proof of Proposition 2}

To reach the GP based formula for \eqref{User Association}, we should have minimization over the objective function, i.e., $$\min_{\boldsymbol{\beta}(t_1)}\sum_{m\in \mathcal{M}}\sum_{g\in \mathcal{G}}\sum_{n_g\in\mathcal{N}_g}\sum_{k\in\mathcal{K}}-\beta_{m,k,n_g}R_{m,k,n_g}(\textbf{P}(t)).$$ Clearly, we have negative elements on the objective function similar to our general formulation in \eqref{CGP_general}. To meet the positive conditions of objective function in GP, we consider $\Xi_1 \gg 1$ and rewrite objective function as $$\Xi_1-\sum_{m\in\mathcal{M}}\sum_{g\in\mathcal{G}}\sum_{n_g\in\mathcal{N}_g}\sum_{k\in \mathcal{K}}{\beta_{m,k,n_g}(t_1){R_{m,k,n_g}(\textbf{P}(t))}}$$ which is always positive. Then, considering a positive auxiliary variable $x_0$, and rewrite the objective function with this new auxiliary variables 
\begin{align}\label{6} \nonumber
& \frac{\Xi_1}{x_0+\sum_{m\in \mathcal{M}}\sum_{g\in\mathcal{G}}\sum_{n_g\in\mathcal{N}_g}\sum_{k\in \mathcal{K}}{\beta_{m,k,n_g}(t_1){R_{m,k,n_g}(\textbf{P}(t))}}} \\ 
& \leq 1,
\end{align}
Now, \eqref{6} can be rewritten as the product of monomial functions based on the AGMA from Section III. B as%
\begin{align}\label{722}
& \Xi_1\left[\frac{x_0}{c_0(t_1)}\right]^{c_0(t_1)}\prod_{m\in{\mathcal{M}},g\in{\mathcal{G}},n_g\in{\mathcal{N}_g},k\in{\mathcal{K}}} \\ \nonumber
& \left[\frac{{\beta_{m,k,n_g}(t_1){R_{m,k,n_g}(\textbf{P}(t))}}}{c_{m,k,n_g(t_1)}} \right]^{c_{m,k,n_g}(t_1)} \leq 1, 
\end{align}
where $c_{m,k,n_g}(t_1)$ and $c_0(t_1)$ are updated from \eqref{7} and \eqref{C0}, respectively. Therefore, the total optimization problem can be transformed into \eqref{beta_eqn}.


\end{document}